\documentclass[preprint,12pt,authoryear]{elsarticle}
\usepackage{amssymb}
\usepackage{amsmath}
\usepackage{todonotes}

\usepackage{hyperref}

\usepackage{soul}

\usepackage{enumerate}
\usepackage[inline,shortlabels]{enumitem}

\hypersetup{
  breaklinks   = true, 
  colorlinks   = true,    % Colours links instead of ugly boxes
  urlcolor     = blue,    % Colour for external hyperlinks
  linkcolor    = blue,    % Colour of internal links
  citecolor    = gray      % Colour of citations
}

\newcounter{JGHCommentsCounter}

\newcommand\anolrevised[1] {{#1}}
\newcommand\wnukrevised[1] {{#1}}

\journal{Journal of Systems and Software}

\begin{document}

\begin{frontmatter}



\title{Reducing Friction in Cloud Migration of Services} %% Article title


\author[Ericsson,BTH]{Anders Sundelin} 
%% Author name
\ead{anders.sundelin@bth.se}
\author[BTH]{Javier Gonzalez-Huerta} 
%% Author name
\ead{jgh@bth.se}
\author[BTH]{Krzysztof Wnuk} 
%% Author name
\ead{krw@bth.se}

\affiliation[Ericsson]{organization={Ericsson AB},
            postcode={371~41}, 
            city={Karlskrona},
            country={Sweden}}
\affiliation[BTH]{organization={Department of Software Engineering},%Department and Organization
            addressline={Blekinge University of Technology},
            postcode={371~79}, 
            city={Karlskrona},
            country={Sweden}}

%% Abstract
\begin{abstract}
Public cloud services are integral to modern software development, offering scalability and flexibility to organizations.
Based on customer requests, a large product development organization considered migrating the microservice-based product deployments of a large customer to a public cloud provider.

We conducted an exploratory single-case study, utilizing quantitative and qualitative data analysis to understand how and why deployment costs would change when transitioning the product from a private to a public cloud environment while preserving the software architecture.
We also isolated the major factors driving the changes in deployment costs.

We found that switching to the customer-chosen public cloud provider would increase costs by up to 50\%, even when sharing some resources between deployments, and limiting the use of expensive cloud services such as security log analyzers.
A large part of the cost was related to the sizing and license costs of the existing relational database, which was running on Virtual Machines in the cloud.
We also found that existing system integrators, using the product via its API, were likely to use the product inefficiently, in many cases causing at least 10\% more load to the system than needed.

From a deployment cost perspective, successful migration to a public cloud requires considering the entire system architecture, including services like relational databases, value-added cloud services, and enabled product features.
Our study highlights the importance of leveraging end-to-end usage data to assess and manage these cost drivers effectively, especially in environments with elastic costs, such as public cloud deployments.
\end{abstract}

%You are encouraged to provide a graphical abstract at submission.
%
%The graphical abstract should summarize the contents of your article in a concise, pictorial form which is designed to capture the attention of a wide readership. A graphical abstract will help draw more attention to your online article and support readers in digesting your research. Some guidelines:
%
%    Submit your graphical abstract as a separate file in the online submission system.
%
%    Ensure the image is a minimum of 531 x 1328 pixels (h x w) or proportionally more and is readable at a size of 5 x 13 cm using a regular screen resolution of 96 dpi.
%
%    Our preferred file types for graphical abstracts are TIFF, EPS, PDF or MS Office files.

%\includegraphics{example.png}

%You are required to provide article highlights at submission.
%
%Highlights are a short collection of bullet points that should capture the novel results of your research as well as any new methods used during your study. Highlights will help increase the discoverability of your article via search engines. Some guidelines:
%
%    Submit highlights as a separate editable file in the online submission system with the word "highlights" included in the file name.
%
%    Highlights should consist of 3 to 5 bullet points, each a maximum of 85 characters, including spaces.
\begin{highlights}
  \item Existing cloud cost calculation models do not take into account newer, value-added, cloud services, provided by the cloud providers.
  \item A large part of the cost of public cloud deployment can be attributed to services that were not originally intended to be deployed in the cloud, such as classical relational database systems, and their license costs.
  \item Real product usage data can be utilized to find optimizations, tracking down ``over-usage,'' inefficiencies and enforce budgets on API users.
\end{highlights}

%% Keywords
\begin{keyword}

  data-driven \sep usage data \sep microservices \sep cloud deployment



\end{keyword}

\end{frontmatter}


\section{Introduction}\label{sections:intro}

Developing large software products and services brings many challenges to software development organizations.
These challenges include
\begin{enumerate*}[label={(\roman*)}]
  \item splitting the product into behaviorally isolated components, allowing individual teams to be in charge of their maintenance and evolution (e.g., by using Service-Oriented Architecture or microservices), and
  \item provisioning just enough compute and storage resources in time to give clients acceptable quality of service.
\end{enumerate*}    
In the early 2010s, the eruption of the cloud paradigm~\citep{armbrust2009above}, with its promise of elasticity and the ability to scale software and hardware systems to match the customers' actual demands, caused many development organizations to consider deploying their software solutions to the cloud.

However, migrating large monoliths to cloud-deployed microservices can pose challenges and risks \citep{jamshidi2013cloud,zhou2023revisiting}.
\citet{taibi2017processes}~ found that 38\% of interview participants manage migration risks by stepwise migration, whereby new or changed features are introduced as separate microservices around the existing monolith, which is gradually replaced. 
This is sometimes referred to as ``the Strangler Fig Pattern'' \citep{fowler2024}.
The migrated software could then be deployed on virtual machines running on bought hardware, functioning like a private cloud.
\anolrevised{In later transformation steps, the organization could migrate services to a public cloud provider, where resources are bought on demand.}

Deploying on a public cloud might require adapting the design of some services to the characteristics of such a platform.
A service that executes on virtual machines running on owned compute hardware might tolerate some inefficiencies as long as the total demand for the service does not exceed the existing hardware's capacity.
However, migrating such a service to a pay-per-use cloud provider, where the organization pays per used virtual CPU core and possibly per database access, will expose these inefficiencies as unnecessary costs.
If brought to the extreme, changing deployment options might impact the whole business case for the service~\citep{tak2012cloudy}. 

In today's highly competitive software market, successful organizations cannot risk taking decisions solely based on expert opinions but need to systematically collect and process data from their systems\footnote{Such data collection and processing must comply both with relevant legislation and existing customer terms and conditions.} to make informed decisions related to their processes and products~\citep{bosch2016Speed}.
Reverse design~\citep{hou2020} is one way in which usage data can inform decisions about the design and architecture of software solutions.

\wnukrevised{This paper presents a case study that examines how deployment cost efficiency changes when migrating a large-scale software product to a different deployment platform.}
The \anolrevised{Product Development Organization (PDO) and a large customer} considered migrating product deployments to a public cloud service provider, and wanted to understand the economic consequences of such a migration.
To assess these consequences, the organization launched a proof-of-concept project where the complete service (including operations and maintenance functions) was migrated to a large public cloud provider.

The customer expected to reduce deployment costs, as reported by~\citet{villamizar2015,villamizar2016}, but the project results contradicted these assumptions.
To understand why this was the case and suggest possible remedies, we analyzed $90$ days of product usage data and interviewed six project members with different roles.
We validated our findings through a focus group \anolrevised{with eight participants having different roles} in the organization.
To \wnukrevised{the best of our knowledge}, this paper is the first to systematically describe the cost implications of migrating a large legacy product to a public cloud provider.

The remainder of the paper is structured as follows:
\anolrevised{in Section}~2, we report the background and related work, and in \anolrevised{Section}~3, we describe the case and our research methodology.
\anolrevised{Section}~4 contains our results, and in \anolrevised{Section}~5, we discuss these.
Finally, in \anolrevised{Section}~6, we conclude the paper and outline future work.

\section{Background and Related Work}

\begin{figure}
  \centering
  \includegraphics[width=0.8\columnwidth]{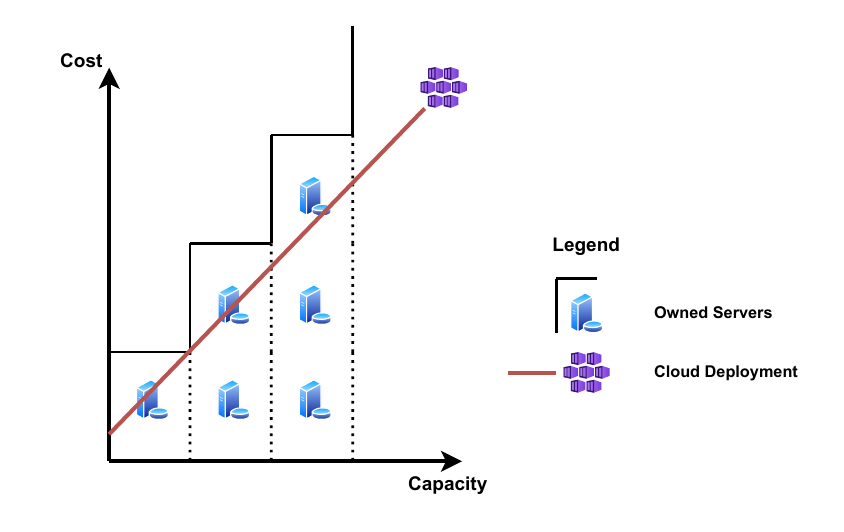}
  \caption{The theory behind cloud economy ~\cite{armbrust2009above}: With owned hardware, users always pay for ``the next server'', no matter how little extra capacity you need (black, zig-zag line), and peak usage will determine the entire system cost.
    With a cloud solution, you should be able to ``pay only what you actually use'' (red line).}
  \label{fig_costVsCapacityInCloud}
\end{figure}

Figure~\ref{fig_costVsCapacityInCloud} illustrates the theory behind the cloud economy, as described by~\cite{armbrust2009above}.
Owning dedicated servers (illustrated by the black line) requires an up-front investment in hardware based on projected usage and plans for expansions well in advance (to manage hardware delivery and installation lead times).
\wnukrevised{When using a cloud provider (illustrated by the red line), the hardware and infrastructure work is outsourced to the cloud provider, who would manage these resources for several systems.}
The foundation for sharing resources is the capability to divide large hardware resources (e.g., server-class computers) into smaller, shareable pieces.
The most common strategy relies on virtualization software, which divides one server into many Virtual Machines (VMs).

We follow the terminology of~\cite{armbrust2009above} and use the term \textit{public cloud} to denote compute (and storage) services being made available to the general public via some on-demand subscription service, called \textit{utility computing}.
In contrast, a \textit{private cloud} refers to a data center that might use cloud technologies, though the services are only available to a select set of users, who typically share some organizational belonging (such as departments or subsidiary companies), that own and operate the underlying hardware and network services.
According to \citet{armbrust2009above}, utility computing is preferable to a private cloud in either of the following cases:
\begin{enumerate*}[start=1,label={\roman*)}]
  \item the demand for the service \textit{varies over time}, or
  \item the demand is \textit{unknown in advance}, or
  \item the work can be \textit{highly parallelized}, as is the case when processing independent work batches. 
\end{enumerate*}

 \wnukrevised{The three main drivers that motivate legacy software to cloud migration are maintainability \citep{fritzsch2019microservices, auer2021}, scalability, and
flexibility \citep{hasan2023legacy}.}  
\citet{andrikopoulos2013} defines four different migration types of increasing ambition level:
\begin{enumerate*}[label={(\roman*)}]
  \item replacing individual architectural components (e.g., a database) with a corresponding cloud offering,
  \item partially migrating some of the application functionality to the cloud,
  \item migrating the application's whole software stack, with the architecture intact (sometimes referred to as ``lift-and-shift''), and
  \item rewriting the application as a composition of services running in the cloud.
\end{enumerate*}
The "lift-and-shift" migration strategy does not benefit the user much, as it is hard to scale, and because of the comparatively long startup time of large VMs \citep{leymann2017native}. 

\wnukrevised{\cite{chen2011cloud} outlines two cloud usage scenarios: the ``unified client'' model, where applications are accessed by a single customer or a small group, and the ``multi-client'' model, where diverse third parties access shared resources, similar to public content distribution. They suggest that for ``unified clients,'' outsourcing is cost-effective when computation outweighs communication. The multi-client setting is generally better for mid-sized enterprises due to the high cost of networking.

Many companies that migrate complex legacy systems prefer to rewrite them using current technologies over splitting up existing code bases \citep{fritzsch2019microservices}. 
To take advantage of cloud deployment, applications should have the following five properties: Isolated state, Distribution, Elasticity, Automated management, and Loose coupling (IDEAL) ~\citep{fehling2014patterns}.}

\wnukrevised{The financial and business aspects of cloud migration were studied by several authors.
Among them,} \citet{walker2009real} compared the cost of purchasing a cluster with the cost of leasing CPU time and storage capacity from an open market by considering the aggregated cost of the resources from a Net Present Value (NPV) perspective, taken over the expected lifetime of the cluster (typically 3-5 years).
\citet{tak2012cloudy} also used the NPV approach but also considers networking costs, arguing that cloud migration is appealing for small or stagnant businesses.
They found that partitioning components is expensive due to the increased networking costs, and that a mix of in-house and cloud deployment can be helpful for certain applications (e.g., to scale out during high traffic).

\wnukrevised{ \cite{konstantinou2012public} compared the Total Cost of Ownership (TCO) of a private cloud (StratusLab) with that of a commercial provider. Their findings indicate that even small-scale private clouds can become more cost-effective within 2–3 years, well before hardware reaches its end-of-life. They also highlight that compute-intensive applications are more economical when organizations maintain full control over the virtualization layer. 
\citet{villamizar2015,villamizar2016} report significant infrastructure cost reductions when transitioning from monolithic architectures to microservices deployed on public cloud platforms like AWS Lambda. 
By implementing three versions of a reader and a writer service, they demonstrate cost savings of up to 77\%, measured per million requests.}

Various approaches have been proposed for auto-scaling cloud resources.
\cite{mao2010cloud} describes how cloud providers enable users to define triggers to scale up or down the cloud resources based on some defined metric (such as CPU utilization, disk or bandwidth usage, etc.)
\wnukrevised{They postulate using simulations to assess cloud budgets properly.}
\cite{boza2017reserved} describe how to use a set of $M(t)/M/*$ queuing theory models to predict load (hence also the cloud budget) for a system over time.
In such a model, arrivals are governed by a Poisson process where the arrivals at time $t$ are a function of $t$: $\lambda(t)$.
With reasonable resolution (e.g., hours), this allows for modeling situations where requests vary by the time of day, as is often the case for services that interact with end users in the same time zones.
In another study, \cite{zhu2010resource} represents the system behavior via an auto-regressive-moving-average with exogenous inputs (ARMAX) model, \wnukrevised{supporting both a fixed time limit and a resource budget for the task at hand.}
They develop a controller that uses the ARMAX model and a Proportional-Integral (PI) control, combined with a reinforcement learning component to minimize costs, subject to time and budget constraints, achieving 200\% improvement over a static provisioning scheme, with a controller overhead of less than 10\%.

Cloud cost calculations~\citep{walker2009real,armbrust2009above} and cloud migration of services~\citep{fritzsch2019microservices,hasan2023legacy} have been researched before. However, \wnukrevised{to our knowledge, no study has analyzed product usage data} when migrating a large (millions of lines of code) legacy system with millions of end-users to a public cloud provider.
Likewise, we have not seen any studies on the actual bottom-line impact of autoscaling services in such legacy systems.
Thus, we believe this paper can provide feedback to the research community on the problems faced by practitioners tasked to migrate existing large-scale software systems to the big public cloud providers.

\section{Research Methodology}
 
\noindent \anolrevised{In this paper,} we study the potential consequences of a strategic decision to migrate a large, existing product offering to a particular public cloud provider.
\anolrevised{To this end, we formulate} the following research questions:

\begin{enumerate}[start=1,label={\bfseries (RQ\arabic*):},wide = 0pt, leftmargin = 3em]
  \item How does the projected cost for the used system resources change when changing \wnukrevised{deployment} from owned hardware to public cloud?
  \item What factors drive the change in cost, and what is their estimated influence?
  \item To what extent can we use service usage data to explain resource usage \wnukrevised{in deployments}?
\end{enumerate}

To answer these research questions, we conducted an exploratory single-case study following the guidelines by \cite{runeson2012casestudy}.
The goal of the case study, described using the GQM framework~\citep{Basily1994GQM} is:
To analyze the usage of the (fully integrated) software product by its end-users \textbf{\textit{from the perspective of}} the organization developing the software product, \textbf{\textit{with the objective}} of understanding the impact of its architecture on deployment cost \textbf{\textit{in the context of}} a potential public cloud migration.

\subsection{The Case and Unit of Analysis}

\begin{figure}
  \centering
  \includegraphics[width=\columnwidth]{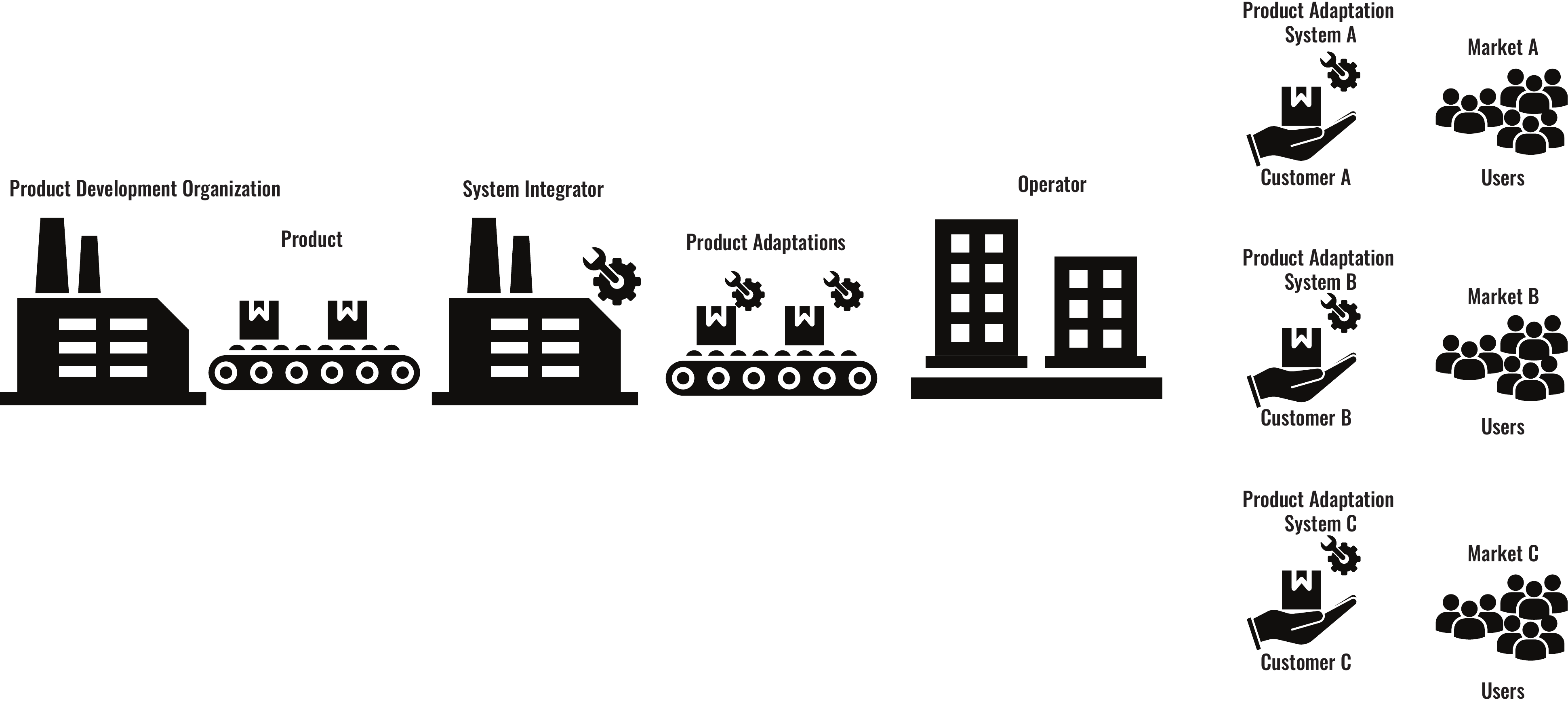}
  \caption{Schematic view of the product and its stakeholders. 
  The Product is developed by the Product Development Organization, customized by System Integrators, made available by Operator(s) who own the product, and realized as installed Systems A, B, C, etc., which are used by Users in different Markets. In our case, the individual Systems are operated by different Customers (A, B, C, etc.), who are owned by a common Operator.}
  \label{fig_case}
\end{figure}

The case under analysis, shown in Figure~\ref{fig_case}, is a global FinTech software product developed by a Product Development Organization (PDO) that exposes financial services via an Application Programming Interface (API).
The product contains some core functions, such as Graphical User Interfaces, but the API allows System Integrators to build end-user Use Cases and adaptations unique to each product installation.
At present, the Product Development Organization sells the product as an integrated software-hardware stack, where each Customer operating the service \wnukrevised{(Operator)} owns the whole solution, including hardware, reselling the financial services to end-users (Users), who typically pay agreed fees (depending on the Use Case) for the usage of the system.
The Users are individual consumers, small merchants, and large companies, such as banks. 
In each market, the product is adapted and customized by System Integrators, companies outside of the PDO, that configure and build the complete end-to-end Use Cases, via the Software Product API and their own code.

\anolrevised{Due to a change of strategy, both the Product Development Organization and the largest Operator wanted to change from the capital-heavy ``owning'' model~\citep{walker2009real} to a more elastic deployment, using a public cloud provider.}
To evaluate the consequences of such a shift, the Product Development Organization developed a deployment solution using a public cloud provider selected by the Operator.
Following the conclusion of the evaluation project, we conducted this case study to elicit recommendations for future adjustments to the software product and the organizational setup of the cloud economy.

\begin{figure}
  \centering
  \includegraphics[width=\columnwidth]{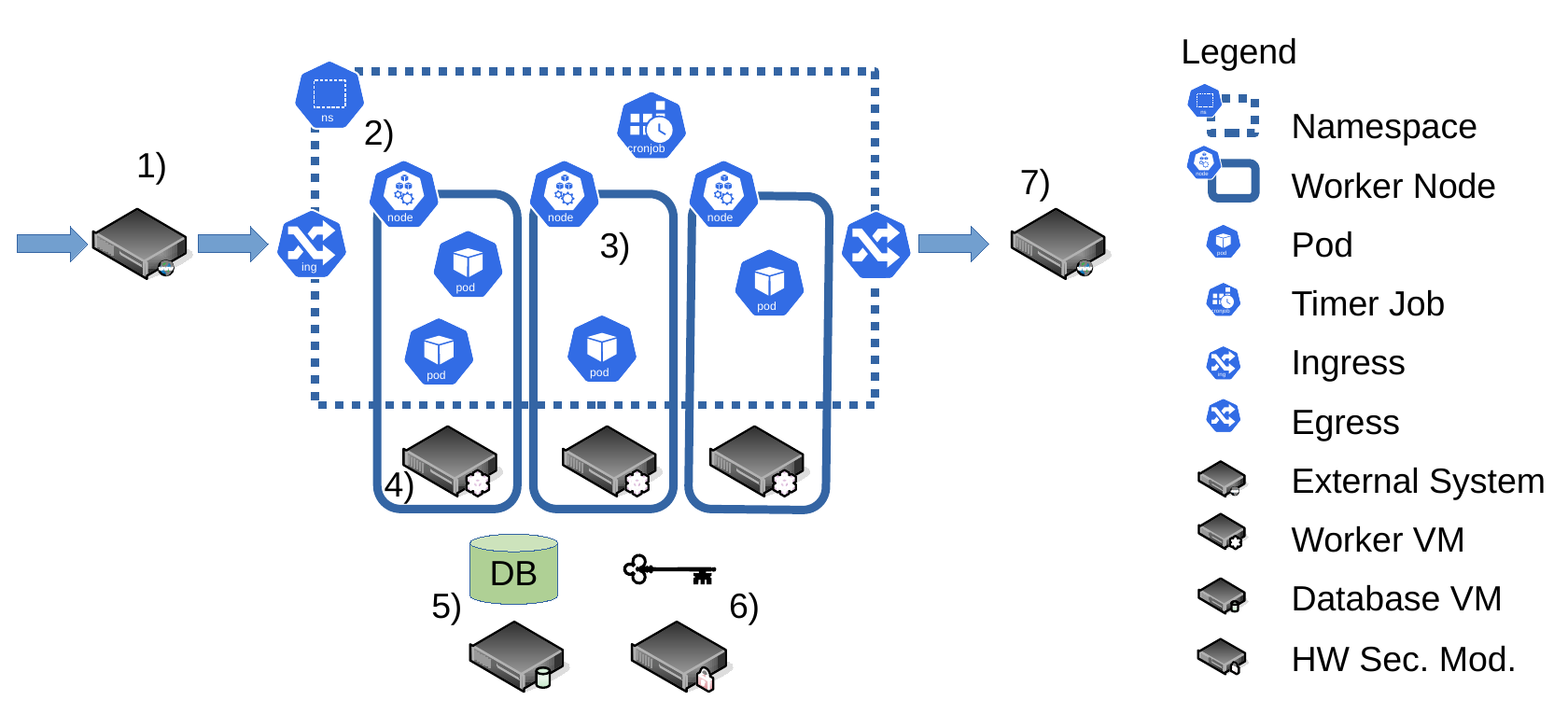}
  \caption{High-level deployment view mixed with Component and Connector view of the current product. Incoming requests~(1) are processed by services running as pods inside a Kubernetes namespace~(2). Pods execute on worker nodes~(3), which are realized via Virtual Machines~(VMs, 4) on top of owned hardware. \anolrevised{Outside the namespace, the database service~(5) runs on separate VMs, and cryptographic keys are stored in a Hardware Security Module~(HSM, 6). Some requests cause interactions} with external systems~(7).}
  \label{fig_system}
\end{figure}

Figure~\ref{fig_system} shows a high-level picture of the product.
A request~(1) is sent to a Kubernetes service running in a namespace~(2).
In the namespace, each service is implemented by one or more pods running on worker nodes~(3).
A minimal configuration contains about 20 Java-based microservices (composed of $\approx3$~MLOC), plus over 80 third-party services.
Each worker node is realized as a Virtual Machine~(VM, 4), running on top of owned physical hardware, using static configuration (e.g., memory and vCPU \anolrevised{characteristics}).
Some application services, such as the audit log and the main application relational database~(RDBMS), are also running as separate VMs~(5) outside the Kubernetes namespace.
\anolrevised{A separate, physical, Hardware Security Module (HSM,~6) manages and protects the cryptographic keys used by the product.
In a few special use cases, the application service communicates with an external system~(7) before returning a response.}
Timers trigger some use cases (such as reports or maintenance tasks), but most services are initiated by incoming requests from external actors.

In production systems, the product is deployed redundantly, where each microservice exists in multiple copies within the same namespace.
\wnukrevised{For higher availability, each deployment might be replicated in multiple namespaces, interconnected via public or private networks, running on geographically separated hardware.}

Overall, the architecture of the system follows the following structure:
\begin{enumerate}[start=1,label={\roman*)}]
\item The system exposes operations (services) of different types (320 in total), where each operation performs a well-defined task.
Most operations are accessed through an HTTP REST API~\citep{fielding2000architectural}.
\item Some operations access services external to the system, waiting for these services to respond before returning a response.
\item Generally, external communication occurs at well-defined places, from a defined subset of operations, meaning that incoming and outgoing requests can be correlated using their operation types.
\end{enumerate}

The product has existed for over ten years and was first deployed, \wnukrevised{as a monolith}, on Linux, running on dedicated physical computers (referred to as ``bare metal'' deployment).
Over time, as the product has grown in capabilities, size, complexity, and number of micro-services, its architecture has evolved, first by adapting to virtualization (running on VMs), then containerization, by adopting the Kubernetes deployment style, as advocated by~\cite{leymann2017native}.
Currently, it is deployed in over 20 different markets across several continents. 

\wnukrevised{The context of this case study can be summarized as:
\begin{enumerate*}[label={(\roman*)}]
  \item the Operator initiating the change to public cloud deployment deployed the product in several markets (countries), having sub-companies (Customer A, B, C, etc.) in each country.
  \item the Operator unilaterally selected deployment to a particular public cloud provider,
  \item the existing Operator deployments already shared resources (such as hardware, but also some expensive services), and
  \item the product kept its original architecture, based on an industry-standard relational database (RDBMS), which scaled only vertically by adding more CPU, memory, and storage to an existing physical or virtual computer.
\end{enumerate*}
}

\subsection{Planning}

We selected the case based on convenience and availability due to the existing partnerships between the studied organization and Blekinge Institute of Technology.

\subsubsection{Methods for Data Collection}

We employed a mixed-methods case study, collecting both quantitative and qualitative data.
In the quantitative part, we employed archival analysis~\citep{Stol2018ABC}, collecting documents and budgets from the migration project and \wnukrevised{90 days of aggregated} usage data from two of the largest system deployments.
In the qualitative part, we employed semi-structured interviews with different roles in the development organization and a focus group to validate our conclusions and give feedback to the studied organization.
We used investigator triangulation when conducting and analyzing interviews and focus groups, and data triangulation to confirm conclusions from a qualitative and a quantitative angle.

\begin{table}
  \caption{Interviewees, how many years they have worked (in general and in the product), and overall job description relevant to the cloud migration project.}
  \begin{tabular}{ l|c|c|l }
     & \multicolumn{2}{|l|}{\textbf{Experience (y)}} &  \\
    \textbf{Role} & \textbf{Work} & \textbf{Product} & \textbf{Major tasks} \\
    \hline
 Cloud Architect & 17 & 12 & \anolrevised{Conceptualizing,} dimensioning \\ 
 Cloud Developer & 16 & 13 & Developing cloud solution \\ 
 \anolrevised{Operations Eng.} & 18 &  8 & Implementing cloud solution \\
 Project Manager & 27 & 14 & Managing people resources \\
 \anolrevised{Requirements Eng.}& 30 & 5 & Eliciting requirements \\
 Sales Architect & 12 & 12 & Technical customer interactions \\
  \end{tabular}
  \label{tab_interviewees}
\end{table}

We interviewed \anolrevised{six PDO employees working in different roles}, selected using convenience sampling.
Details about the participants are available in Table~\ref{tab_interviewees}.
Through time-boxed (30-minute) interviews, we tried to understand
\begin{enumerate*}[start=1,label={(\roman*)}]
\item the drivers and the context of the migration to the public cloud provider,
\item their subjective experiences during the project, and
\item their perceptions about using usage data to assess the deployment strategy of the system.
\end{enumerate*}

Having explained our data anonymization and reporting protocol, we recorded and transcribed the interviews using Microsoft Teams\textsuperscript{\copyright}.
The automatic transcriptions were then revised and anonymized by one of the researchers in the research team, and the resulting transcription was presented to the interviewee for feedback and possible clarifications.
The anonymized transcripts are available upon request.

\wnukrevised{To validate our findings}, we also conducted a focus group session with eight participants (including roles such as Technical Sales, System Architect, Cloud Architect, Developer, Tester, and Operations/DevOps Engineer), who commented on our conclusions.
The focus group was also recorded, transcribed, and anonymized.

\subsubsection{Methods for Data Analysis}

The data analysis has been iterative, although some activities have also been intertwined.
Quantitative data was analyzed using visualizations, descriptive statistics, and correlation and regression models.
Qualitative data has been analyzed using open coding~\citep{saldana2016}, and transcripts and coding scheme are available upon request.
As an example, the quote ``\textit{It was not so clear what they were trying to solve [with the move to cloud deployment].}'' was coded with [Weak and unclear requirements], which, together with the code [Limits of lift-and-shift], also was applied to the quote (from a different respondent): ``\textit{We were just thinking of how we can run [the product] in the cloud. We did not think about the impact~[\dots] in terms of cost or in terms of operational efficiency.}''

\subsection{Generalizability}

Although this paper contains data and findings from a particular system developed and deployed by specific organizations, we have strived to describe the surrounding context in adequate detail while still respecting the need for confidentiality required by commercial enterprises.
As~\citet{flyvbjerg2006} states, cases play an essential role in human learning, and it is, in fact, possible to learn from a single case.

The characteristics and context we describe are far from unique, and we hope the case description enables others to judge whether our findings apply to other systems.
However, we cannot claim generalizability across all possible organizations, systems, or cloud providers, although we believe the study brings interesting insights and lessons for practitioners and researchers.

\section{Results}

\subsection{Projected cost change due to change of deployment platform (RQ1)}

\wnukrevised{The first public cloud estimate turned out to be four times more expensive than the existing private cloud due to four key reasons}:
\begin{enumerate*}[label={(\roman*)}]
  \item using a single estimate (for the smallest deployment) as the base of the cost estimation for the entire Operator group of companies,
  \item the lack of cost requirements in the initial specification,
  \item the lack of relevant feedback loops---i.e., the requirement to come up with the complete five-year cost before ever deploying anything to the cloud, and
  \item the failure to consider that some---crucial, but lightly used---cloud services could be shared among all Operator companies.
\end{enumerate*}

 \anolrevised{The Project Manager emphasized the intent to replace existing Open-Source bundled services (e.g., log visualization, intrusion detection and other services not directly tied to product-specific features) with commercial cloud services}: ``\textit{If it is in the cloud, as a service from the Cloud Provider, or in the Marketplace, then we will use that service. \anolrevised{[\dots] OK, so then we can finally get a nice security monitoring tool, instead of Tripwire}}.''

\anolrevised{The Cloud Developer} stated the lack of feedback as a prime reason for the increased cost estimate: ``\textit{When you deploy a solution to the cloud, it takes a while to come to the optimal solution[\dots] You can never get an optimal level on paper[\dots] You need some time to get some feedback from live data to make an application.}''
He also stated the lack of feedback as a hindrance to choosing suitable external services: ``\textit{Log analytics is responsible for collecting all the logs from the infrastructure. It is a very expensive service in the Cloud Provider. Before I have tested with a real live installation, how can I determine how much storage I should use for it?}''

\subsubsection*{Sharing Used Services}
Once the initial cost estimate was realized, the Sales Architect started to get more involved: ``\textit{My first reaction was, no one has looked into this architecture from the point of optimization. I went into challenging the teams[\dots] `OK, why do we need to ingest all that data into [the Log Analytics Service] in Cloud-Provider'}?''
This intervention reduced the estimated deployment cost considerably---from four times (300\%), down to 50\% extra cost, compared to the existing ``private cloud'' solution.

Thus, while the initial cost estimate was unrealistic, even after optimizations suggested by the Sales Architect were considered, the deployment cost for the system in the public cloud would still be 50\% higher than the existing private cloud costs.

\subsection{Factors influencing public cloud deployment costs (RQ2)}

Based on feedback from the Cloud Developer and the Operations team, we also looked into how the required resources varied over time to estimate whether the used resources could be sized accordingly.

\subsubsection*{Autoscaling Worker Nodes}

\begin{figure}
  \centering
  \includegraphics[width=\columnwidth]{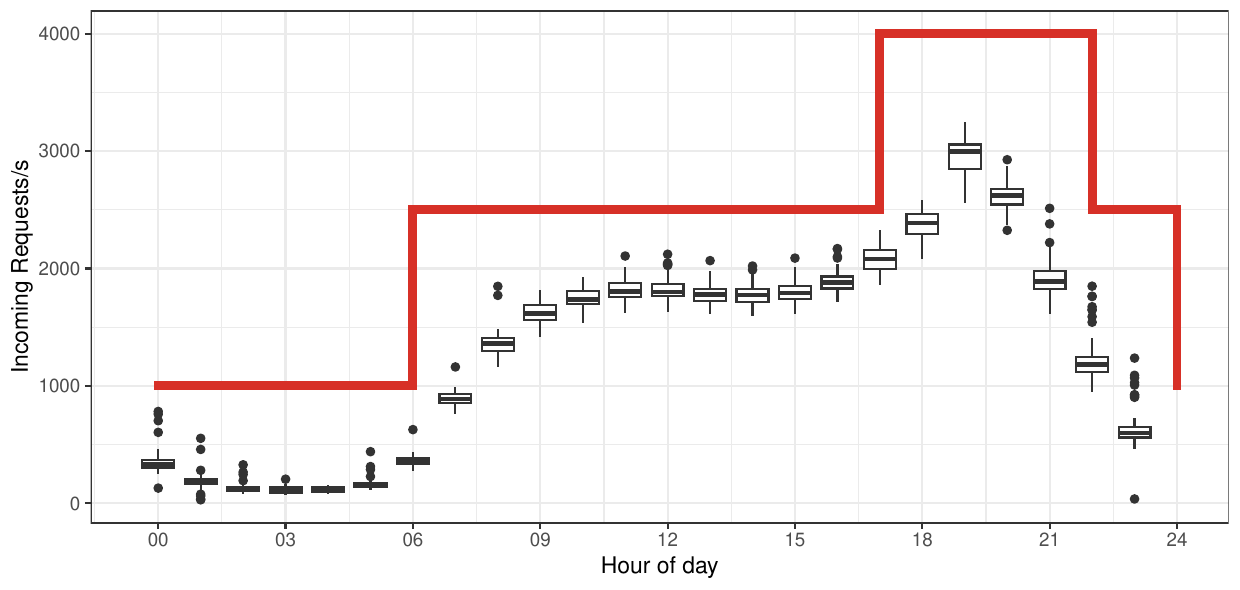}
  \caption{A grouped boxplot showing how the number of incoming requests per second varies by the hour of day, measured during 81 days in late 2023. Daytime traffic is flat, followed by an evening peak and drop-off after midnight. The red line indicates a possible adaptation of needed resources to the incoming load.}
  \label{fig_average_tps}
\end{figure}

Figure~\ref{fig_average_tps} \wnukrevised{shows, using a grouped boxplot, how the number of incoming requests per second varies throughout the day for a representative deployment.}
The figure shows a flat daytime load, followed by a stable peak, starting around 17:00 and receding around 21:00, which reaches about two-thirds (median $\approx\frac{3000}{1800}$) more than the daytime load.
 \wnukrevised{During night-time, incoming requests are much lower than during daytime, with traffic increasing around 06:00. The figure also shows a tentative adjustment to the needed resources.}
The observed daily variability is consistent with feedback from the Cloud Developer: ``\textit{At night, we hardly have any traffic. During the day, we have plenty of traffic, but we are paying for the same resources day and night.}''

\wnukrevised{Based on the existing traffic pattern, the Cloud Developer suggested adding auto-scaling to the product to adjust the deployment size and needed resources according to incoming load.}
Converting the red line in Figure~\ref{fig_average_tps} into needed resources (assuming the maximum resources required corresponds to the peak usage---i.e., 4000 requests per second), finds that, based on this traffic pattern, the required resources over time can be expressed as a fraction of the peak resource usage, as per expression~\ref{eqn_usage}.

\begin{equation}
\frac{6\cdot\frac{1000}{4000}+11\cdot\frac{2500}{4000}+5+2*\frac{2500}{4000}}{24} \approx 61\%
\label{eqn_usage}
\end{equation}

As the dimensioning factor is the peak resource usage, in a private-cloud scenario---where the hardware cost is fixed---it makes little sense to reduce resources via autoscaling.

The chosen public cloud provider charges resource usage per started VM, regardless of the vCPU load, meaning that the number of Kubernetes worker nodes are the only resources that could be autoscaled.
For redundancy reasons, all deployments require at least two worker nodes.

\subsubsection*{Cloud Cost Break-Down}

\begin{figure}
  \centering
  \includegraphics[width=1 \columnwidth]{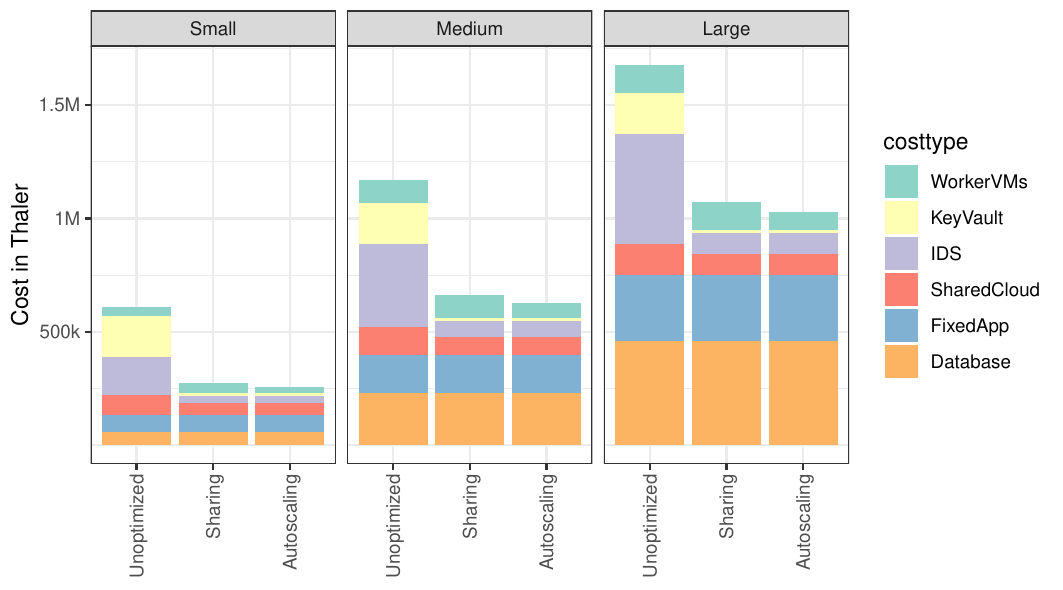}
  \caption{
    \anolrevised{Estimated costs, per cost type and deployment size and type. Costs are expressed in the fictional \textit{Thaler} currency, having a fixed, hidden, exchange rate to USD. \\
    Cost types are:~\textbf{WorkerVMs}~the Kubernetes worker nodes (Virtual Machines);
    \textbf{KeyVault}~the Cloud Provider Key Management solution for secure key storage;
    \textbf{IDS}~the Cloud Provider AI-based Intrusion Detection System;
    \textbf{SharedCloud}~services potentially shared across deployments (e.g., firewalls, observability tools, syslog infrastructure, and the associated network connectivity);
    \textbf{FixedApp}~services needed by each deployment (e.g., storage, backup, network connectivity); and
    \textbf{Database}~license and support costs.}
  }
  \label{fig_cost_comparison}
\end{figure}

\anolrevised{We analyzed the Bill of Materials\footnote{Anonymized data found in: \url{https://tinyurl.com/Sundelin-Cloud-Migration}} (BoM) used by the project, which specifies contents and their costs (initial investment plus five-year total license and maintenance costs) for Small, Medium, and Large deployments. 
Costs for the three alternative solutions are estimated for each deployment size:
\begin{description}
  \item[Unoptimized] refers to the initial estimate, where each deployment is separate, with no shared resources. This solution also heavily uses the AI-based log analytics solution from the Cloud Provider.
  \item[Sharing] refers to the estimate in which some expensive cloud resources are shared among the deployments, and logs sent to the AI-based log analytics solution were trimmed.
  \item[Autoscaling] contains all improvements from the \textbf{Sharing} solutions, but also adds dynamic scaling of the Virtual Machines used as Kubernetes worker nodes.
\end{description}
The original BoM specified the costs in US dollars.
However, for business confidentiality reasons, we converted the costs to a fictional currency, called the \textit{Thaler} ($\mathcal{T}$), which converts to USD via a fixed and hidden exchange rate.

Figure~\ref{fig_cost_comparison} contains the cost per cost type for the three deployment sizes and solutions. 
The \textit{Unoptimized} costs grow from $\approx609~\text{k}\mathcal{T}$ for the Small deployment, to $\approx1.68~\text{M}\mathcal{T}$ for the \textit{Large}.
The figure also shows the dominant cost of the Intrusion Detection System (IDS) and the Key Vault for all three deployment sizes.
The relative database cost grows with the deployment size, more than doubling from Small ($\approx 10\%$) to Large ($> 25\%$) deployments.
}

\begin{table}
  \centering
  \caption{\anolrevised{Total and worker node cost for three deployment variants in the fictional \textit{Thaler} currency, and the savings realized by adding autoscaling, as a percentage of the total \textbf{Sharing} cost.}}
  \begin{tabular}{l|r|r|r|r|r}
     & \multicolumn{2}{|c}{\textbf{Total cost in $k\mathcal{T}$}} & \multicolumn{2}{|c|}{\textbf{Worker node cost}} & \textbf{Relative} \\
    \textbf{Size} & \textbf{Sharing} & \textbf{Autoscaling} & \textbf{Sharing} & \textbf{Autoscaling} & \textbf{Savings}\\
    \hline
    Small  & 272  & 257 & 42 & 26 & $5.8\%$ \\
    Medium & 664  & 627 & 102 & 65 & $5.5\%$ \\
    Large  & 1070 & 1026 & 122 & 77 & $4.2\%$ \\    
  \end{tabular}
  \label{tab_workernode_cost_ratios}
\end{table}

\anolrevised{Table~\ref{tab_workernode_cost_ratios} contains the total and the worker node costs expressed in \textit{Thaler}, as well as the expected savings realized when adding \textit{Autoscaling} to the \textit{Sharing} deployment variant.
The table shows that we could expect total savings of $\approx4-6\%$, if worker nodes were scaled according to the traffic pattern shown in Figure~\ref{fig_average_tps}.
}
The Cloud Architect expected more savings to materialize from autoscaling: ``\textit{We need to work more on that concept of autoscaling, and actually use exactly what you need at this moment.}''.

\begin{figure}
  \centering
  \includegraphics[width=\columnwidth]{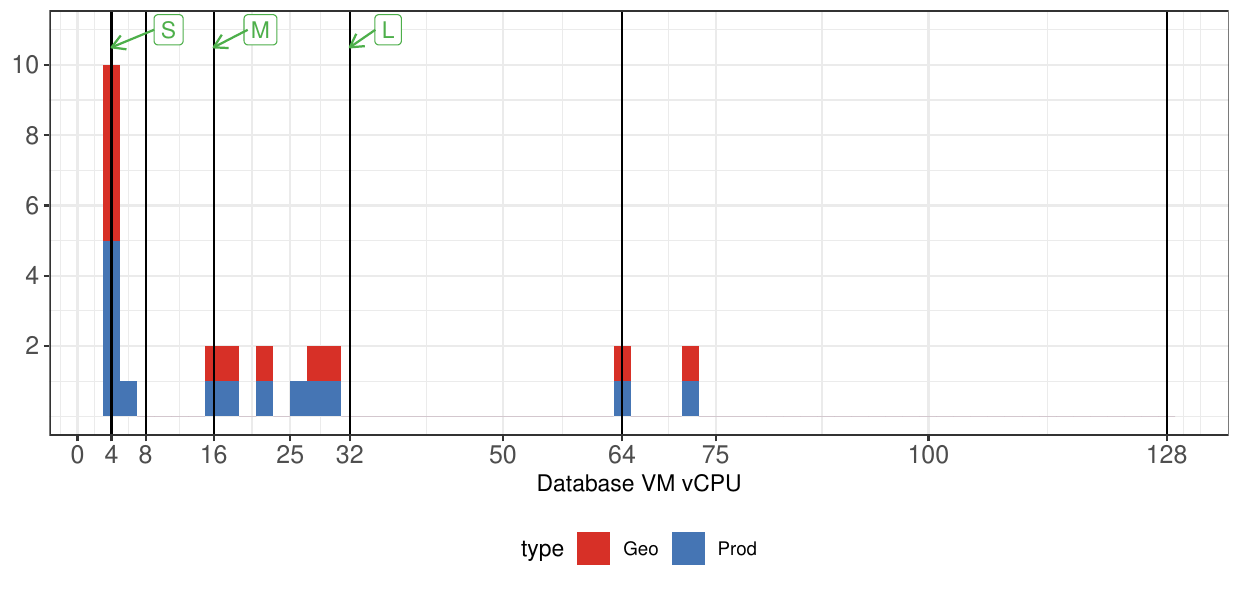}
  \caption{Number of deployments with a particular number of vCPUs in the database Virtual Machines. 
  Five deployments have 4 vCPUs, and the biggest deployment has 72.
  All but two deployments are geographically redundant.
  The sizes of the database VM in Small, Medium, and Large deployments are marked with S, M, and L.
  The available public cloud VM sizes are marked with black vertical lines.}
  \label{fig_db_vcpu}
\end{figure}

\anolrevised{In contrast, the database service, running on statically configured Virtual Machines (VMs), forms a much larger part of the overall cost, and its importance increases as the system grows.
The reasons for the cost growth are both that the Cloud Provider provides VMs of fixed, exponentially spaced sizes (i.e., 4, 8, 16,\dots vCPU), and that the license costs for the database are proportional to the number of available virtual CPUs on the database VMs.}

Figure~\ref{fig_db_vcpu} shows the size of the database VMs used by some deployments (both production and geographically redundant systems).
The five smallest systems require only four vCPUs, but the largest requires 72, implying that the corresponding available public cloud VM would need 128 vCPUs.
Running the database on such a VM would increase license costs by 78\% for the largest deployment.
Taken together, should all these systems deploy on fixed-size public cloud VMs (whose available sizes are indicated by the vertical lines in the figure), the license costs would increase by 31\%.
The exponential sizing of available VMs implies avoiding database expansions will significantly impact operating costs, especially for larger systems.

\anolrevised{We discussed the database solution and its cost with our interviewees, and it is clear that:
\begin{enumerate*}[start=1,label=(\roman*)]
    \item the product requires a particular brand of Database;
    \item while the chosen Cloud Vendor provides other databases ``as-a-Service,'' none of these databases are fully compatible with the required brand: ``\textit{No one is using Database in Cloud-Provider. Everyone is using other databases like SQL Server or Postgres, which are available natively there.}'';
    \item using a third-party Database service would introduce unwanted dependencies on other vendors (e.g., in troubleshooting, fault isolation, and other contractual obligations);
    \item the Cloud-Provider recommends using Virtual Machines to provide the required Database: ``\textit{They don't have Database `as-a-Service.' But they have like a recipe how to do it, and then you have to do it yourself.}''
\end{enumerate*}
}

\subsection{Using service data to explain resource usage~(RQ3)}

Given that the database service constituted a significant factor driving public cloud deployment costs, we decided to investigate \textit{how} the product was being used for two of the major deployments.
With accurate usage data, it would be possible to pinpoint which parts of the product contributed to the cost and relate these to the revenue-generating parts.

\subsubsection*{Budgeting Resources}

\begin{figure}
  \centering
  \includegraphics[width=\columnwidth]{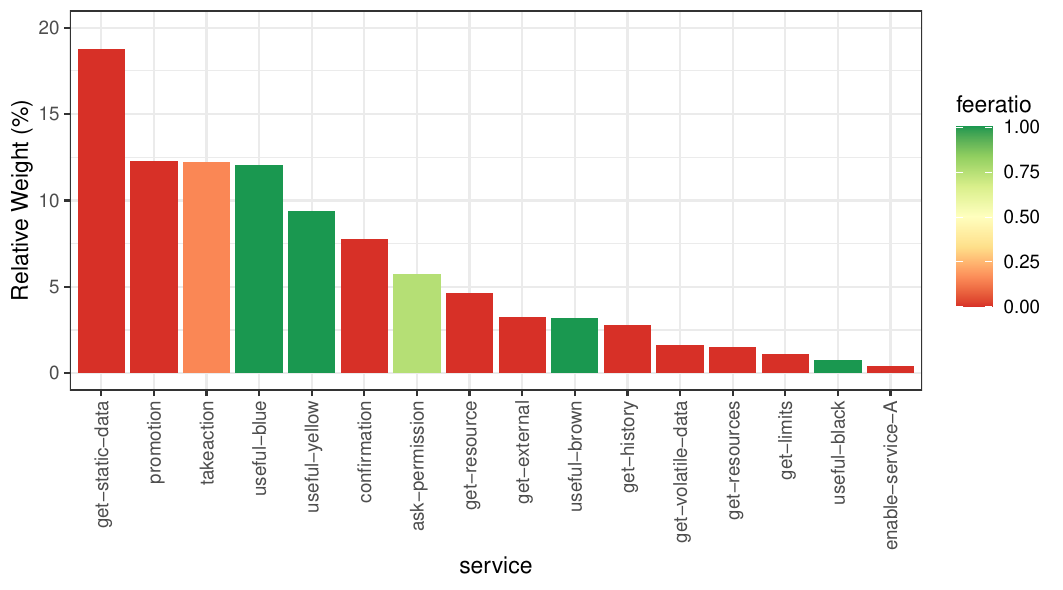}
  \caption{Relative resource weight of the top 16 (out of 320) different services being used, comprising 97.5\% of the used resources. The color indicates the extent to which the service was generating revenue.}
  \label{fig_adjusted_relative_weight}
\end{figure}

To assess resource usage by the different services, we define the metric $\mathcal{W}[S]$, called \textit{weight} for service \textit{S}, calculated over some time of interest:

\begin{equation}
  \mathcal{W}[S] = t_i[S] \cdot n_i[S] - t_o[S] \cdot n_o[S]
  \label{eqn_weight}
\end{equation}

Using equation~\ref{eqn_weight}, we define $\mathcal{W}[S]$ in terms of: $t_i[S]$, the average service time (latency) for service $S$; $n_i[S]$, the number of service invocations; $t_o[S]$, the average latency of external services used by service $S$; and $n_o[S]$, the number of requests to external services ($\geq0$) made by service~$S$.
Importantly, in this model, a database request is considered an internal service, as the database runs as an integrated part of the system.

\wnukrevised{The current product allows clients unlimited requests without considering the cost of each used service.}
Figure~\ref{fig_adjusted_relative_weight}   \wnukrevised{shows the most resource-intensive services in the system, ranked by relative weight in relation to the combined weight of all services.}
\wnukrevised{The figure shows that only four of the top 16 services are directly generating revenue.}
The most used service (``get-static-data'') is reading static data, which would have been easy to cache.

The services ``takeaction'' and ``askpermission''  \wnukrevised{only partially generate revenue.}
In ``takeaction''  \wnukrevised{about 87.5\% of requests are for free, while }``askpermission''  \wnukrevised{requests may be declined by end users}.
In total, services using about 30\% of resources generate direct revenue, considering that two services only partially do so.
Based on usage data analysis of two systems, we find that a moderately sized cache would reduce the number of requests to the most used service ``get-static-data'' by between 10\% and 50\%, which corresponds to a 2\%-10\% reduction of overall resource usage.

\subsubsection*{Focusing on Database Statements}

With moderate effort, it is possible to look into what statements the product is executing towards the database, and given application traces and source code, these statements can be attributed to services.

\begin{figure}
  \centering
  \includegraphics[width=\columnwidth]{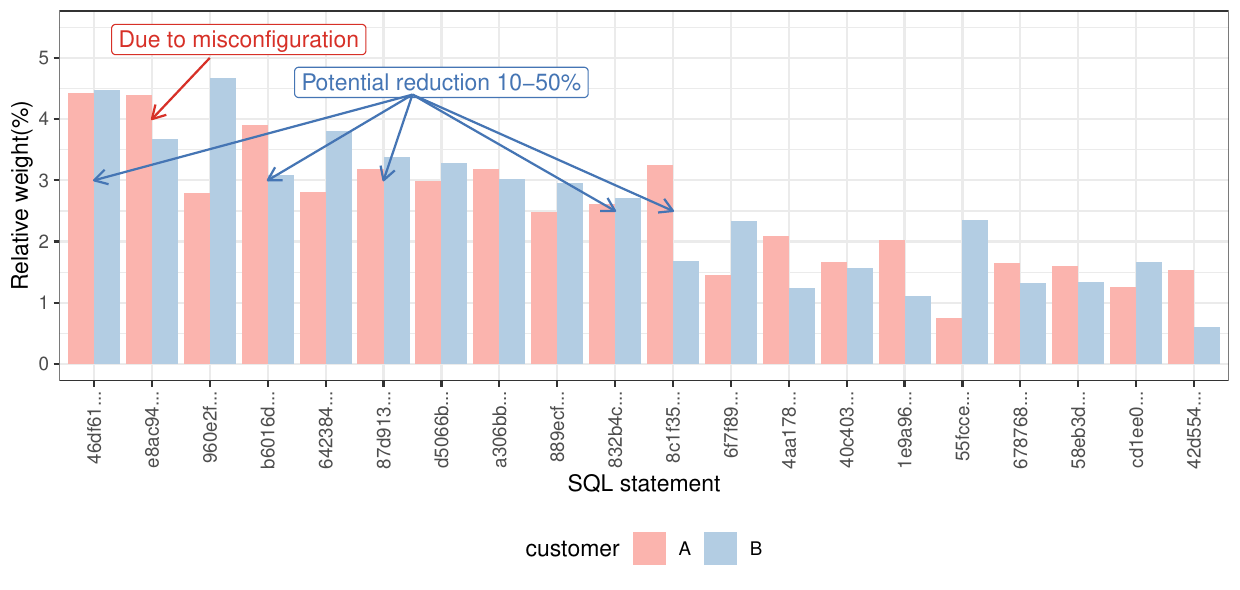}
  \caption{\anolrevised{Relative weights for the top SQL statements for the two largest deployments. A misconfiguration (an unused, partially enabled feature) causes the second heaviest statement, and five of the eleven heaviest could have been reduced by 10\%-50\% if ``get-static-data'' calls had been cached. Statements are arranged by average relative weight, and the graph includes statements having over 1.5\% weight in either A or B.}}
  \label{fig_top_peak_sql}
\end{figure}

Figure~\ref{fig_top_peak_sql} shows the top \anolrevised{20 SQL statements for the two largest product deployments. 
\wnukrevised{We found five statements related to the above-mentioned ``get-static-data'' service, and could be cut} between 10\% and 50\% if caching had been used.
The second most used statement was related to a feature neither customer used, but the product assumed to be active due to misconfiguration.}

\subsubsection*{Summary---Usage Data to Explain Resource Usage (RQ3)}

We can summarize the results with a quote from the Sales Architect: ``\textit{[Solution Integrators] look at our API specification and their focus is to get the functionality [the Use Cases] to work. Their focus is not to develop something performance-oriented. I have never seen someone questioning `this design, is it efficient from a performance perspective?'}''
As shown in Figure~\ref{fig_db_vcpu}, this approach was adequate as long as the system was executing on fixed, owned, and paid-for hardware.
Selecting the chosen Cloud Provider, with its fixed-size VMs, shows that the costs become painfully obvious, at least for the larger systems.

\anolrevised{Figure~\ref{fig_adjusted_relative_weight} shows that only six of the top 16 used services generate revenue, and that the most used service, consuming about $\approx 18\%$ of the database resources, is reading mostly static data, which would have been easy to cache. 
We see in Figure~\ref{fig_top_peak_sql} that this service affects five of the most costly database statements, and this figure also reveals that the second most resource-intensive statement, using between $3.5-4\%$ of the database capacity, was entirely unnecessary, caused by misconfiguration of the product.}

We see that usage data is critical to understanding current ``hot spots'' and highlighting inefficiencies that would go unnoticed.
Our respondents agree with this statement: ``\textit{You need real-life usage data to determine the chips, the size, and which services to use}.''

\section{Discussion}

  Unlike previous studies \citep{villamizar2015,villamizar2016}, our study considered moving a complete application stack, including security, operations, and maintenance services, from an existing owned hardware platform to a large public cloud provider.
  \wnukrevised{The product had evolved over ten years from ``bare metal'' deployment to containerization with Kubernetes, but still relied on a classical, monolithic, relational database backend, with no plans to change this architecture.}

\noindent \textbf{How would deployment cost change when moving to a public cloud?}
We found that, for the initial estimate, specific cloud provider \textit{value-added services}, such as a custom Log Analysis service and a Key Vault (managing cryptographic keys for the application), together with the assumption that these were required for each deployment caused the cost to increase by 300\%, as a five-year total, compared to the existing deployment.
Even if these assumptions were relaxed by sharing services across deployments and reducing the volume of logs processed by the Log Analysis service, we found that the deployment cost in the public cloud would exceed the existing deployment cost by 50\%.

\noindent \textbf{What factors caused this increase in cost?}
  We found that the existing relational database was the major cause of the projected increase in cost when deploying to the chosen cloud provider.
Had contracts been renegotiated, another Cloud Provider been chosen, or the database replaced by an Open-Source alternative, this cost driver might have been reduced or disappeared.

\wnukrevised{We found from interview and usage data that the product's workload was heavily time-dependent, with predictable evening peaks, flat daytime load, and very low nighttime load. 
Due to the Cloud Provider's fixed pricing model based on VM size regardless of vCPU load, changing the number of Kubernetes worker nodes was the only way to scale automatically. 
However, such scaling would yield comparatively small savings of around 5\%.}

Significant cost savings materialized when our Sales Architect questioned whether all requested Cloud Provider services were needed and whether they provided enough value to match their cost.
This is a classic requirements engineering problem: although a service might appear extremely useful, its value must exceed the cost it incurs to make economic sense. 

Clearly, the Cloud Providers know how to price and market their services to users of their systems.
In our case, it was obvious to the Sales Architect that some Cloud Provider services could be shared across deployments and that the new, expensive, AI-based Intrusion Detection System should be used with care,
\wnukrevised{and only be fed data where it could efficiently use its AI capabilities, such as standard system and access logs, but not product-specific application log files. }

Larger---but more complex to implement---savings would have been realized by being more careful when accessing product services and, indirectly, the database.
For small systems (4-vCPU VMs), the database license and support costs represent about 10\%, but for larger systems using 32 vCPUs, these correspond to over 25\% of the deployment cost.
In a cloud scenario, trying to avoid increasing the database VM size as long as possible makes sense, which is precisely the opposite of the linearly scalable situation depicted in Figure~\ref{fig_costVsCapacityInCloud}.
\anolrevised{Our respondents found it counterintuitive that autoscaling represented small savings relative to the database and shared services costs.}

\noindent \textbf{How can we use service usage data to explain resource usage?}
We found that service usage data of incoming and outgoing requests and database usage statistics are crucial to finding the parts of the system that drive the resource usage and, therefore, the costs.
Using such data, we quickly found several potential improvements that would reduce peak time load on the database by $10-20\%$, which translates to substantial savings for the larger systems (e.g., up to $10$~fewer vCPUs).
Importantly, these savings also apply---albeit to a lesser degree---in the private cloud scenario, as reduced database license costs.

\noindent \textbf{Summary:}
Enforcing budgets is a common way to make users aware of economic constraints.
With the system currently deployed in a private cloud, no constraints were imposed on system integrators using the product API.
This led to a situation where a significant portion (between 10\%-50\%) of calls to the most used service were unnecessary and could have been cached in an appropriate place, with substantial savings in resource utilization.
Likewise, we found an unwanted feature that was partly enabled, causing at least 4\% unnecessary database load.
In a public cloud scenario, the importance of caching and performance evaluations of features will increase.

Given our traffic pattern, autoscaling would generate marginal cost savings at best.
\wnukrevised{The most significant cost savings would come from} carefully choosing cloud provider services, reducing unnecessary API calls, and reviewing the used product features.
We also found that API users, such as system integrators, seem to focus more on functionality than on measuring the performance implications of their implemented use case.
The product development organization could add budgeting tools based on live system data, allowing customers and system integrators to assess and predict the resource usage of individual use cases.
This would allow customers to detect and react to unnecessary API calls by system integrators building frequently used use cases.
\wnukrevised{Furthermore, such budgets would allow end-to-end use case performance testing to verify compliance with the allocated resource budget.}

Although complex autoscaling models, such as those envisioned by \cite{boza2017reserved} and \cite{zhu2010resource}, might apply to this system, other factors, such as use case optimization and database scalability, play a more significant role than these models.

\section{Conclusions and Further Work}

The early works on cloud computing \citep{armbrust2009above, walker2009real} envisioned that computing in a public cloud would be used as a utility, and researchers such as~\citet{villamizar2015,villamizar2016} have claimed deployment cost savings of up to 77\%.
We find that the opposite would be the case for a real-life, non-trivial system serving millions of users via hundreds of use cases for over ten years.
Even if obvious optimizations were adopted (sharing services across deployments, restricting the use of expensive security services), the deployment cost would increase by around 50\% if the suggested public cloud provider were used and the system architecture were unchanged.
This suggests that the IDEAL properties of~\cite{fehling2014patterns} need to apply to \textit{all} relevant product components and services for the benefits to materialize.

Similar to findings by \citet{taibi2017processes}, our product had evolved using the ``Strangler Fig Pattern'' \citep{fowler2024}, whereby the number of microservices had increased to around 100, of which about 80 were open-source tools.
However, most microservices still used a commercial relational database (separating the storage via private schemas), whose license costs were directly tied to the number of virtual CPUs used for the database Virtual Machines.
At a minimum, the product or service owner needs to be prepared to buy a ``Database-as-a-Service'' or ensure that the chosen cloud provider supports Virtual Machines (VMs) of adequate size granularity.
Compared to the TOSCA vision of~\cite{leymann2017native}, old relational databases are still heavily reliant on VMs, as the database products have accumulated many features over the years.

In our study, the cloud provider's new, advanced, AI-based Intrusion Detection System represented a large part of the projected cost.
\wnukrevised{Care had to be taken to keep costs down so as} not to overload it with data that was unlikely to yield much benefit for the system's security.
Sharing costs across deployments also lowered the public cloud deployment cost, but could not match the existing private cloud costs.
Such new commercial services, or sharing costs across deployments, are not considered in existing models of cloud cost effectiveness~\citep{chen2011cloud}, \wnukrevised{suggesting a need for new, updated models considering} the new cloud deployment offerings.

Regarding auto-scaling~\citep{mao2010cloud}, simple or more complex algorithms, in the vein of~\citet{boza2017reserved} and~\citet{zhu2010resource}, might be effective to some extent, but in our case, even though the peak-hour traffic exceeded the median daytime load with $\approx66\%$, and night-time load was even lower, the net savings would be marginal, around 5\%.

More savings would be realized using budgeting tools when describing API usage to system integrations to avoid unoptimized use cases.
However, these savings would also apply to private cloud deployments, in which case they would postpone expansions of the existing deployment (i.e., reducing the need for additional server hardware).
This implies that we agree with the conclusions of~\cite{konstantinou2012public}, at least for systems like ours that keep their original architecture when migrating.
This adds context to the studies of migrating legacy systems to the cloud~\citep{chen2011cloud, hasan2023legacy}, and suggests a potential necessary condition (i.e., cloud-aligning the architecture), before it is economically viable to migrate to a public cloud provider.
We find that the data collected by~\citet{auer2021} contain some of the metrics (e.g., response time, number of requests per time unit) that can be used to build budgeting tools, provided that these metrics are measured on a level granular enough to assess which services are driving the resource usage.

Our research suggests that the economic benefits of migrating a large legacy system to the present-day commercial public cloud providers are highly dependent on:
\begin{enumerate*}[label={(\roman*)}]
    \item the architecture of the migrated system---in particular, how dependent it is on traditional relational database systems,
    \item the use of \textit{additional value-added} services provided by the cloud provider---in particular, services used for operation and maintenance (including security monitoring) of the migrated system and
    \item to what extent it is being ``over-used'' by clients who might not realize that they are utilizing the system more than needed---the marginal cost of an extraneous invocation leading to an additional required vCPU will be immediately visible in the monthly bill from a public cloud provider. In contrast, owned hardware is expanded much \wnukrevised{less often}; therefore, the cost is much more obscured.
\end{enumerate*}

We intend to continue studying the impact of visualizations and budgeting tools on how system integrators can realize the actual deployment cost of their solutions, regardless of whether the system is deployed in a public or private cloud scenario, and invite others to do the same.

\section*{Declarations}
\noindent Mr Sundelin declares that although he has been employed by Ericsson AB during the writing of this study, neither this company nor any other would gain or lose financially from reporting this research.
The other authors declare that they have no known competing financial interests or personal relationships that could have appeared to influence the work reported in this paper.

\section*{Acknowledgements}
\noindent This research was supported by the KKS Foundation through the KKS SERT Research Profile project (Ref. 2018010).


\bibliographystyle{elsarticle-harv} 
\bibliography{references}

\end{document}